\documentclass{IEEEtran}
\IEEEoverridecommandlockouts
\usepackage{amsmath, amssymb, amsfonts}
\usepackage{algorithmic, graphicx, textcomp, xcolor}
\usepackage{tabularx}
\usepackage{url}
\usepackage{array}
\graphicspath{{FIGURESprotectionstats/}}
\usepackage{calc}  
\usepackage{enumitem}  

\newcommand{\poneone}{\raisebox{-1.7ex}{\includegraphics[scale=0.6]{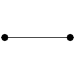}}~}
\newcommand{\ptwooneone}{\raisebox{0ex}
{\includegraphics[height=7pt]{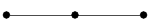}~}}

\begin{document}

\title{The statistical spread of transmission outages on a fast protection time scale based on utility data 
}

\author{{Ian Dobson, D. Adrian Maldonado, Mihai Anitescu 
\\Iowa State University and Argonne National Laboratory, May 2025
\vspace{-0.1cm}
}
\thanks{I. Dobson is with Electrical and Computer Engineering Department,
Iowa State University,
Ames IA, USA; email:
dobson@iastate.edu. D.A. Maldonado and M. Anitescu are with Argonne National Laboratory.  This work was supported in part by the Office of Electricity, Advanced Grid Modeling (AGM) program, U.S. Department of Energy, under contract DE-AC02-06CH11357 and by NSF grant 2153163 and PSerc project S110.}
}

\maketitle

\begin{abstract}
When there is a fault, the protection system
automatically removes one or more transmission lines on
a fast time scale of less than one minute. The outaged 
lines form a pattern in the transmission network. 
We extract these patterns from utility outage data, determine
some key statistics of these patterns, and then show how
to generate new patterns consistent with these statistics. The
generated patterns provide a new and easily feasible way to model the overall effect
of the protection system at the scale of a large transmission system.
This new data-driven generative modeling of protection is expected to contribute to simulations 
of disturbances in large grids so that they can better quantify the risk of blackouts. 
Analysis of the pattern sizes suggests 
an index
that describes how much outages spread in the transmission network at the fast timescale.

\end{abstract}


\section{Introduction}

\looseness=-1
We present observed
utility outage data that shows patterns in line outages at the protection system time scale of less than one minute. The line outages occurring during each minute mostly appear on the network as single lines or small tree subnetworks.
The protection system routinely limits the outages to simple patterns, most frequently to outages involving one or two lines. However, much rarer but more impactful patterns with many line outages also appear in the data. 

We extract and analyze such patterns from historical line outage data for two transmission systems. 
We compute key statistics of these observed patterns that describe how outages propagate on the network at the fast protection time scale.
These statistics are then used to calibrate a model capable of generating new patterns consistent with those statistics that can start from any initial line outage.
This amounts to novel data-driven generative modeling of the effects of protection at the transmission system level. 

The new systems-level statistical modeling of fast protection actions is significant for two reasons: First, it provides an alternative to detailed modeling that is based on real data and is relatively easy to apply. This new alternative is promising since it is difficult to model the intricate details of the protection system 
for an entire transmission system; the barriers include access to the detailed protection data for an entire system, 
coordinating protection models and models for transmission system statics and dynamics, and the many ways that the 
protection system can act when operating as intended, 
operating in an unusual network condition, 
or misoperating \cite{DobsonPMAPS18,FlueckPESGM20}.
Second, there is a need to model protection at the transmission system scale: The protection system contributes to the risk of cascading or extreme weather causing blackouts,
so that giving a practical way to model the effects of the protection system at the transmission system scale is useful.
It is also generally worthwhile to extract and characterize the actual overall protection system behavior from utility data at the transmission system scale. This helps to further ground the subject in reality and can help guide the development and validation of detailed models.

Our approach to reproducing observed protection operation patterns is conceptually close to the generative modeling techniques employed in recommendation systems and online discussion threads \cite{Aragon2017}. 
Rather than only obtaining a set of descriptive statistics of the observed data, generative models contrast with descriptive methods by their ability to generate synthetic observations. 

In summary, the main contributions of this paper are:
\begin{itemize}
\item Extract and present observed utility data showing the patterns of how line outages did propagate at the fast protection time scale in two transmission networks.
\item Compute key statistics describing the size and form of the patterns. In particular, we find that the pattern size has a heavy tailed distribution.
\item Show how to generate patterns on the network that match the key statistics. This is a new data-driven statistical model of outage propagation at the fast time scale that applies at the transmission system level. 
\item Assess the data-driven generative model performance in terms of a distance between generated patterns and observed patterns.
\item  Suggest a new index that measures how much outages spread in a transmission system due to protection.
\end{itemize}

\section{Literature review}
\label{literature} 

Models of protection system backup operation and misoperation can be used to identify multiple N--k contingencies to be included in lists of initial contingencies that are used to evaluate transmission system robustness or to initiate cascading failure simulations. 
Indeed, the previous work most nearly related to the overall approach of this paper is Zhou et al. \cite{ZhouPS24}, which finds spatial patterns of initial contingencies in utility data called contingency motifs. Contingency motifs occur much more frequently in practice than multiple outages chosen randomly with equal probability from the utility network. 
Contingency lists using these more frequent patterns are much more effective in capturing the probability of multiple contingencies \cite{ZhouPS24}.
This paper and \cite{ZhouPS24} both analyze patterns in utility data, but the thrust of the analysis is quite different: \cite{ZhouPS24}
takes the patterns as given and analyzes their frequency to find the motifs and improve contingency lists, whereas we estimate key statistics of patterns and use these to generate similar patterns. 
\cite{ZhouPS24} only analyzes patterns initiating cascades and includes disconnected patterns, while we analyze all the connected patterns appearing at the fast time scale.
The smaller patterns that we study in this paper are also frequent enough to be contingency motifs but we also analyze and generate the larger patterns that are too rare to be motifs. 

There is also detailed protection system modeling aimed at improving contingency lists.
Chen and McCalley \cite{ChenPS05} model protection groups and stuck breakers to trace the higher probability series of events online under different substation and maintenance conditions to alert operators to higher probability contingencies.
Yang et al. \cite{YangIREP07} analyze hidden stuck breaker failures in a 24-bus system to find critical contingencies.
 Jiang et al. \cite{JiangNAPS21} use a Markov chain to determine the steady state probabilities of  NERC standard categories of contingencies.  The failure and repair rates are estimated from utility data.

There is extensive literature assessing the cascading risk of large transmission system blackouts by simulating models as surveyed in \cite{CFTFPESGM08,WGPS16}. 
Most of this literature outages lines without any detailed consideration of the protection system. 
However, there are advances in modeling some aspects of the protection system in this context. 
Rios et al. \cite{RiosPS02} 
model protection system misoperation by a constant probability when a fault lies in the vulnerability region of a relay.
Chen et al. \cite{ChenEPES05} model hidden protection failure of a line exposed by a neighboring outage with a probability that changes according to the line loading. 
Yu and Singh \cite{YuPS04} obtain probabilities of failure to operate and the undesired trip from a steady state Markov chain and then use these probabilities to simulate cascades in a 24-bus power system.
Dobson et al. \cite{DobsonPMAPS18} model the protection system in some detail and account for stuck breakers when cosimulating protection and power system dynamics in a 130-bus power system.
Another application where detailed protection modeling has been applied is assessing resilience to the risk of extreme weather. Ciapessoni et al. \cite{CiapessoniSG16} assess this risk in an 80-bus power system, and their models include busbar and double circuit faults and stuck breakers.

 Anders et al.~\cite{AndersPS06}  describes in detail the failure statistics of air blast breakers, including the distribution of time to failure for forced outages. 
    Bollen \cite{Bollen93} describes probabilities of a variety of breaker failures.
    Grant \cite{GrantPhD25} extracts condition-dependent failure probabilities and rates from breaker data. 
    
    Protection actions with multiple outaged lines include common mode events.
     Mittelstadt et al.~\cite{MittelstadtPMAPS04} find the mean time between common mode failures of BPA transmission lines and Keel  et al.~\cite{KeelPESGM12} describe  the statistics of WECC transmission line common mode failures. In particular, Keel  et al.~\cite{KeelPESGM12} confirm that almost all the common mode events in WECC start in the same minute.
     Other data sources for common mode and dependent failures are reviewed in \cite{PapicPESGM14}.
     These common mode failure statistics can be expressed as failure rates and used  in steady state reliability models \cite{BillintonPESGM12, PapicPS17}.
     
Except for \cite{ZhouPS24},  all these studies advance the detailed modeling of specific mechanisms of protection operation and misoperation. 
These studies complement and have a different scope than our investigation of all the outage spreading that actually occurred at the fast time scale.
Detailed modeling of specific mechanisms allows the effect of that mechanism to be assessed in simulations of blackout risk in small transmission system subnetworks, whereas our approach based on observed data faithfully reflects the combined effect of all the mechanisms in large-scale transmission systems.

Generative modeling of general cascading outages is used by Kelly-Gorham et al. \cite{KellyEPSR20,KellyPS24} as part of modeling transmission system resilience in the CRISP Computing Resilience of Infrastructure Simulation Platform. 
Starting from a random initial line outage, successive line outages are chosen according to the observed distribution of resilience event sizes and the observed distribution of distances between successive line outages in resilience events.
The line and generator restoration times are also sampled from their observed lognormal distributions.
Similar generative modeling is used by Cheng et al. \cite{ChengES22} to sample network cascading caused by an earthquake.

The generative modeling of line outages in CRISP and  \cite{ChengES22} has a similar overall objective to this paper: to generate outages consistent with the statistical patterns of observed events. 
However, it models all the cascading and outage processes combined and does not separately consider protection processes at the fast time scale. 
As this paper shows, fast protection processes have a special structure different from cascading or weather-induced outages in general. 
Most notably, outage propagation at the fast protection time scale produces small connected subnetworks that are mostly trees, whereas general cascading or weather-induced outages propagate locally, as well as more globally to disconnected sets of lines. 
This explains why this paper controls how the new lines are directly attached to the evolving patterns, whereas CRISP and \cite{ChengES22} sample new lines according to their network distance from the initial line outage.

While the only previous generative models we know of in power transmission systems are in \cite{KellyEPSR20,KellyPS24,ChengES22}, generative models of cascading phenomena have been used in other applications.
Generative models have been used to model cascades of product recommendations in social networks and in discussion forums \cite{Leskovec2007, Gomez2011}. 
For example, Leskovec et al. examine the topology of blog posts when some of them become popular and observe that the cascade size distribution presents a heavy tail and follows a Zipf distribution.
They develop a generative model based on contagion dynamics that is able to accurately model the statistics of the observed cascades \cite{Leskovec2007}. 
Gomez et al. introduce a more sophisticated model based on preferential attachment \cite{Gomez2011}. They derive a likelihood function used for parameter estimation and they fit their model to four different blogging communities. 
The model can replicate observed statistics, and the fitted parameters are used to draw conclusions about communication habits. A later review of generative models for online discussion threads points out the successes and diverse uses of these models, such as comparison of discussion platforms, predicting user behavior, and evaluation of platform design \cite{Aragon2017}. 

\section{Utility data}
\label{data} 

The main part of the Bonneville Power Administration (BPA) transmission system is in Washington and Oregon states.
We analyze 19 years of historical outage data recorded by BPA and publicly available at \cite{BPAwebsite}.  
The New York State Independent System Operator (NYISO) transmission system outage records cover New York State and parts of neighboring states and Canadian provinces, with more network detail in New York State. The NYISO outage data is publicly accessed from its website \cite{NYISOwebsite},
 and 12 years of data  
 from 2008 to 2020 
 are processed according to the method in \cite{CarringtonNAPS21}. 

\begin{figure}[ht]
\centering
\includegraphics[width=0.8\columnwidth]{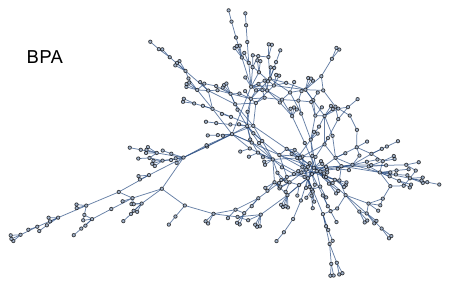}
\vspace{3mm}
\includegraphics[width=0.8\columnwidth]{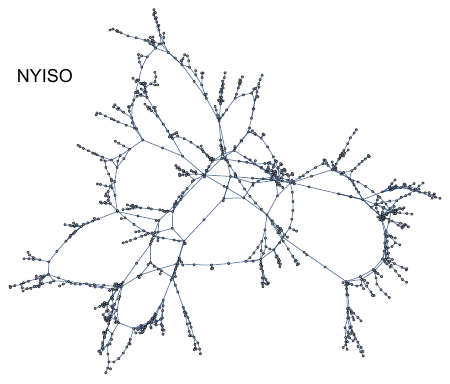}
\caption{Networks for pattern analysis. }
\label{network}
\end{figure}

\looseness=-1 
After data cleaning, mainly standardizing bus names, the power transmission network is deduced from the outage data itself using the method in  \cite{DobsonPS16}, which ensures that the outaged lines can be easily located on the network.
Any isolated portions of the network are removed to ensure a connected network. Outages of lines outside the main component and outages of the Pacific DC intertie in the BPA network are removed.
This yields a network of 608 lines for BPA, including multiple circuits, and 1238 lines for NYISO.  The BPA network can be reduced to a single-line network of 468 lines by combining the multiple circuits into single lines. Fig.~\ref{network} shows the single-line BPA and NYISO networks.
We will find the outage patterns in the single line networks in Fig.~\ref{network} and account for the multiple circuits in the BPA network as a final step.

For both transmission systems, the automatic line outage data include the outaged line specified by its sending and receiving end buses and the line outage times at a one-minute granularity.  Other details of the outage data that are not used in this paper are described in \cite{DobsonPS12} and \cite{CarringtonNAPS21}.
Then, the automatic line outages are grouped together into the lines that start their outage within the time span of the same minute\footnote{Repeats of outages of the same line in the same group are removed.}.
Each group of lines is called a ``generation" of line outages in  \cite{DobsonPS12,DobsonPS16,ZhouPS24},
although these references use a slightly different definition of generation\footnote{References 
\cite{DobsonPS12,DobsonPS16,ZhouPS24} define a generation as groups of outages that are separated by at least 1 minute, giving generations that sometimes include more outages. For example, if outages occur in each of 2 successive minutes, outages in each minute would be a generation in this paper, whereas all these outages would be included in one generation as defined in \cite{DobsonPS12,DobsonPS16,ZhouPS24}.
}.

\begin{figure}[t!]
\centering
\includegraphics[width=0.69\linewidth]{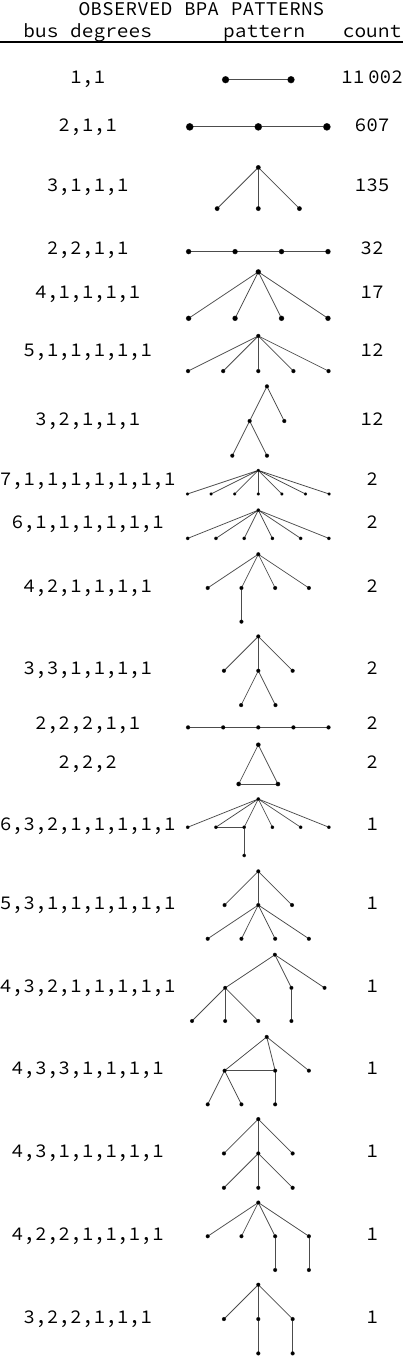}
\caption{\label{patternsBPA}
Degree sequences of patterns and their number observed in BPA data together with an example pattern for each degree sequence.  Note that any multiple circuit outages in the BPA patterns are not shown.}
\end{figure}

Our outage time data have one-minute granularity, so the group is simply the outages with that minute recorded as their start time.
Most of the groups of outages in one minute are a connected subnetwork of the transmission system network, and in that case 
the pattern is that group of outages. For a small fraction of cases  (2.6\% for BPA and 1.1\% for NYISO), there is more than one connected component in the group. 
In that case, 
we consider each of the components in the group as a pattern, as combined failures stemming from fast protection actions can be assumed to result in a connected pattern. 
In summary, the patterns are the connected automatic outages that start at the same minute. It is easy to automatically process the outages to group them into patterns. 

 There are a few external causes that can cause multiple nearby faults 
    at the fast time scale of less than one minute. For example, one of the large patterns observed in the BPA data is caused by an earthquake. 
    Lightning could locally cause multiple faults at the fast time scale, either by multiple ground contacts of a single cloud to ground strike \cite{StallAOT09}, or multiple cloud to ground strikes \cite{PetersonESS24}. 

    Groups of patterns of various types can be further analyzed. For example, the top 6 most common dispatcher cause codes for the BPA outages in patterns with 3 or more outages are: Lightning 15\%,
 Foreign Trouble 15\%,
 RAS Initiated 12\%,
 Unknown 11\%,
 Forced (Configuration) 10\%,
 Terminal Equipment Failure 6\%. 
 If the line physical location is known, allowing weather data to be related to the pattern, then weather causes can be analyzed more reliably than with cause codes.

\begin{figure}[t!]
\centering
\includegraphics[width=0.69\linewidth]{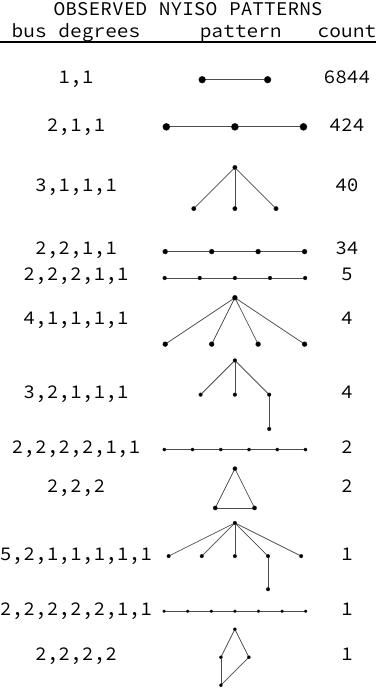}
\caption{\label{patternsNYISO}
Degree sequences of patterns and their number observed in NYISO data together with an example pattern for each degree sequence. Since  the degree sequence determines the pattern for all these observed patterns, all the generated patterns are shown and the count of each degree sequence is the same as the count of each pattern.}
\end{figure}

\section{Observed patterns and their statistics}
\label{statistics} 

Patterns observed in the BPA and NYISO data are shown in 
Figs. \ref{patternsBPA} and \ref{patternsNYISO}, together with their bus degrees and a count of how many times those bus degrees appear in the data.
(Recall that the bus degree is the number of lines incident at each bus, and that the list of bus degrees is known as the degree sequence in network theory.)
Most of the patterns are small trees, with only a few patterns with loops.
For BPA, there are 11836 patterns (623 patterns per year) of which 93\% have one line, 5\% have two lines, and 2\% have 3 or more lines. For NYISO, there are 7362 patterns (609 patterns per year) of which 93\% have one line, 6\% have two lines, and 1\% have 3 or more lines.
 
\begin{figure}[htb]
\includegraphics[width=0.9\linewidth]{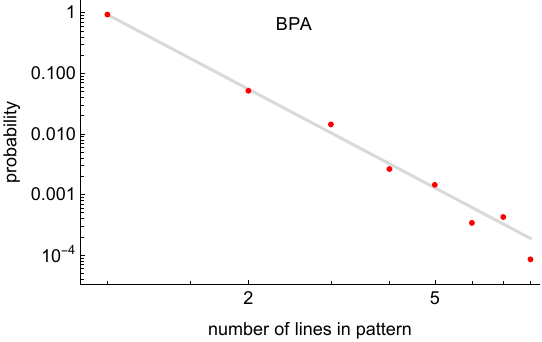}
\\~\\
\includegraphics[width=0.9\linewidth]{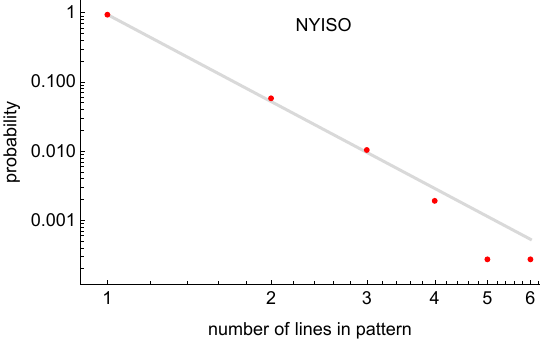}
\caption{\label{distnlinesinpatterns}Empirical distributions of number of lines in patterns and Zipf fits }
\end{figure}

The simpler patterns in Figs. \ref{patternsBPA} and \ref{patternsNYISO} are much more common than the more complicated patterns, many of which only occur once or twice. 
 This is expected in a well-engineered protection system, as it is designed to quickly halt the propagation of outages.
The distributions of the number of lines in the patterns are shown by the red dots in the log-log plots in 
Fig.~\ref{distnlinesinpatterns}.
These observed data in Fig.~\ref{distnlinesinpatterns} can be fit with a Zipf distribution using the maximum likelihood method  in \cite{ClausetSIAMREVIEW09}.
The Zipf distribution (often called the zeta distribution or discrete Pareto distribution) is a  heavy-tailed probability distribution $Z$ on the positive integers with
 \begin{align}
 \hspace{-2mm}
P[Z=k]&=\mbox{probability of $k$ lines in pattern}\notag\\
&=\frac{1}{\zeta(s)}
\frac{1}{k^{s}}, \quad k=1,2,3, ...
\label{zipf}
\end{align}
\looseness=-1
where $\zeta$ is the Riemann zeta function. The discrete probabilities (\ref{zipf}) lie on a 
 straight line of slope $-s$ on a log-log plot.
For Fig.~\ref{distnlinesinpatterns}, $s=4.09$ so that the slope of the line is $-4.09$. 
This is a heavy-tailed distribution for which the probability of a large number of lines in a pattern decreases quite sharply as the number of lines increase, but the decrease is slower than exponential. 
In particular, $s=4.09$ implies that the probability is multiplied by  $2^{-4.09}=0.06$ when the number of lines in the pattern doubles. The Zipf distribution (\ref{zipf})  in effect extrapolates the observed data to allow arbitrarily large patterns; one could alternatively use a truncated Zipf distribution with an upper bound on pattern size.

The observed patterns can be reproduced by starting with a single line out and successively adding lines.
The key statistical features of the patterns describe the number of lines in the pattern and where the lines were attached to form the pattern. 
The number of lines in the pattern should match the Zipf distribution (\ref{zipf}) that fits the distribution of number of lines in the observed patterns. 
While forming the pattern, lines can be attached at buses of degree 1 in the pattern or at buses of degree $\ge2$ in the pattern. 
The probability of lines attached at buses of degree 1 should match the corresponding probability in the observed patterns.

\begin{figure}[htb]
\centering
\includegraphics[width=\columnwidth]{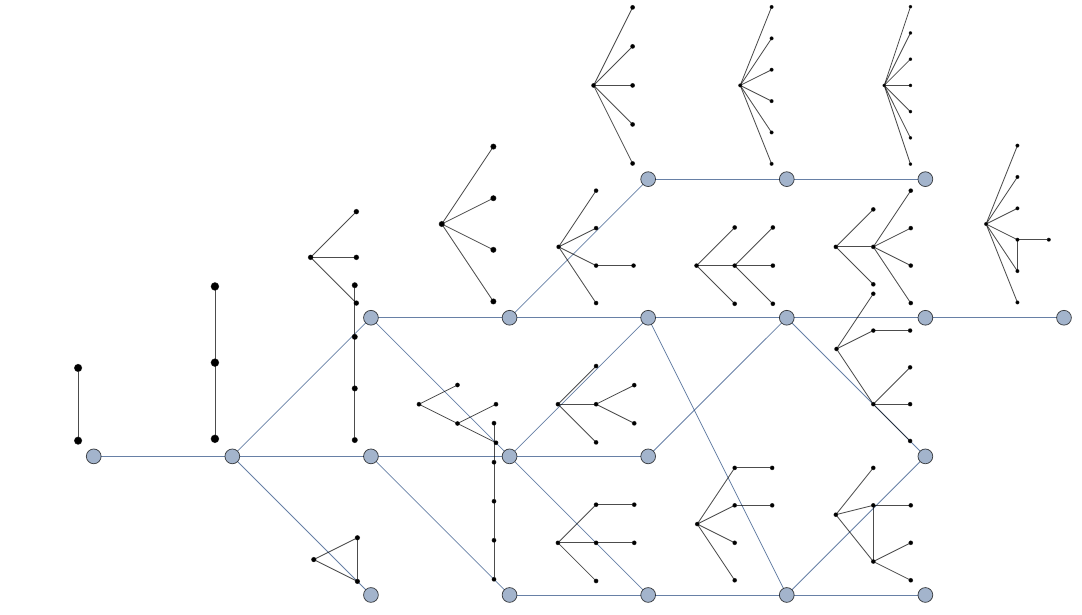}
\caption{Relation of the observed degree sequences of patterns in BPA. Each degree sequence is labeled by an example pattern with that degree sequence.
Degree sequences differing by the addition or subtraction of one line are joined by an edge.}
\label{relationgraphBPA}
\end{figure}

The single line pattern \poneone can add a line at one of its buses to become the pattern \ptwooneone.
Further line additions at buses with degree $\ge2$ starting from \ptwooneone can yield the star patterns such as \raisebox{-0.4em}{\includegraphics[height=15pt]{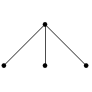}} and \includegraphics[height=9pt]{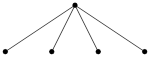}. 
And line additions starting from \ptwooneone at buses of degree 1 can yield linear patterns such as \includegraphics[height=11pt]{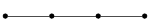} and \includegraphics[height=11pt]{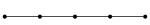}. 
Other patterns such as \raisebox{-0.4em}{\includegraphics[height=15pt]{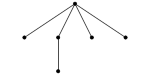} }result from adding lines at buses both of degree $\ge2$ and of degree 1.
A particular pattern can be formed in multiple ways. 
For example, \raisebox{-0.4em}{\includegraphics[height=15pt]{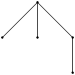}} can be formed by either attaching a line to the degree 2 bus of \ptwooneone followed by a line to a degree 1  bus of \raisebox{-0.4em}{\includegraphics[height=15pt]{p3111}}, or by attaching a line to a degree 1 bus of \ptwooneone followed by a line to a degree 2 bus of \includegraphics[height=11pt]{p2211}. 
The observed ways to successively add a line to obtain the BPA patterns are shown in Fig.~\ref{relationgraphBPA}.

Since each line joins 2 buses, the total number of lines in pattern $\pi$ is 
\begin{align}
n(\pi)= \mbox{(sum of the bus degrees in $\pi$)}/2
\end{align}
For example, \ptwooneone with degree sequence $2,1,1$ has (2+1+1)/2=2 lines.

Where lines have been attached in forming a pattern can be estimated from the degree sequence of the pattern. First, consider the case of a pattern $\pi$, which is a tree. Then, all the patterns that evolved from \poneone to $\pi$ are also trees.
Each addition of a line 
attaches one end at some bus in the evolving pattern, and since all the evolving patterns are trees, the other end is not attached to the evolving pattern. In particular, 
each line addition at a bus of degree 1 changes the corresponding 1 in the degree sequence to 2 and appends a new 1 to the degree sequence.
This bus degree of 2 can be further incremented by the addition of other lines, but these further additions are all to a bus of degree $\ge2$. 
Therefore, the total number of additions of lines at a bus of degree 1 in forming a tree pattern $\pi$ is given by
\begin{align}
    n_{1+}(\pi)&= \mbox{number of bus degrees $\ge2$ in $\pi$} \,.
    \label{n1plustree}
\end{align}

A large majority of the observed patterns are trees.
However,
if the pattern $\pi$ is not a tree, the line additions forming $\pi$ include a line addition that joins two buses of the evolving pattern to form a loop in the pattern.
When counting the number of additions of lines at a bus of degree 1, (\ref{n1plustree}) remains valid if only one of the joined buses has degree 1. However, if both of the joined buses have degree 1, then (\ref{n1plustree}) will count that single line addition twice. For example, according to (\ref{n1plustree}),  $n_{1+}(\includegraphics[height=7pt]{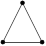})=3$. 
This double counting can be corrected for some simple non-tree patterns by defining 
\begin{align}
    n_{1+}(\includegraphics[height=7pt]{p222.pdf})&=2\mbox{ and }
    n_{1+}(\includegraphics[height=7pt]{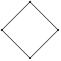})=3
    \label{n1plusloops}
\end{align}
as special cases. This gives an exact answer for the observed NYISO patterns.
However, the situation is more complicated for the more elaborate non-tree patterns in BPA because there are many possible orders in which the pattern can be assembled, and only the orders in which the line forming the loop is attached to join buses of degree 1 in the evolving pattern have double counting. 
Since there are only two more elaborate non-tree patterns for the observed patterns in BPA and none for NYISO, and there is an overcount by one in only some of the possible assembly orders, it is an underestimate that is a good approximation for the purpose of estimating $p_{1+}^{\rm observed}$ in (\ref{poneplusobserved}) below to define 
$n_{1+}$ by (\ref{n1plustree}) together with the special cases (\ref{n1plusloops}).
Note that we do not know from our data the initial line of each observed pattern or the order in which lines were added to obtain each observed pattern. 

Consider all the patterns $P_{\ge3}$ with 3 or more lines, all of which are produced by adding lines to \ptwooneone.

Each pattern $\pi$ in $P_{\ge3}$ added $n(\pi)-2$ lines to \ptwooneone.
Hence there are a total of $N=\sum_{\pi\in P_{\ge3}} (n(\pi)-2)$ line additions to \ptwooneone to produce all the patterns $P_{\ge3}$.

Each pattern $\pi$ in $P_{\ge3}$ has had $n_{1+}(\pi)$ line additions to a bus of degree 1 starting from \poneone and $n_{1+}(\pi)-1$ line additions to a bus of degree 1 starting from \ptwooneone.
Therefore, in producing the patterns $P_{\ge3}$ starting from \ptwooneone, the observed empirical probability of adding a line to a bus of degree 1 is 
\begin{align}
p_{1+}^{\rm observed}=\frac{1}{N}\sum_{\pi\in P_{\ge3}} (n_{1+}(\pi)-1) \,.
\label{poneplusobserved}
\end{align}
Since lines must be added at either a bus of degree 1 or a bus of degree $\ge2$, the empirical probability of adding a line to a bus of degree $\ge2$ in producing $P_{\ge3}$ is $1-p_{1+}^{\rm observed}$.

For patterns observed in BPA, $p_{1+}^{\rm observed}=0.201$, and 
for patterns observed in NYISO, $p_{1+}^{\rm observed}=0.533$.

The BPA network sometimes has multiple circuits joining the same two buses, and we now account for the possibility of a second circuit outaging when there are multiple circuits.
The probability $p_{\ge 2}^{\rm circuits}$ of an additional circuit outaging in a line with multiple circuits is estimated as the number of generations with $\ge\!2$ multiple circuit outages divided by the number of generations with outages of a line with multiple circuits. For the BPA network, $p_{\ge 2}^{\rm circuits}=0.07$.

\section{Generative model of fast protection patterns}
\label{sec:fast_protection_modeling}

This section describes the fast protection modeling of probabilistically and
successively adding lines to an initial single line outage to generate patterns.
We first consider multiple circuits joining the same two buses to be single circuits (i.e. consider the network 
 to be  simple) and account for the possible outages of the other circuits as the last step.
 
 The patterns are formed by starting with a single line out and successively adding lines until the process stops. The process is probabilistic so that 
it can be repeatedly sampled to produce a range of patterns that reproduce some key statistical features of the observed patterns.
This process is only intended to reproduce these key statistical features of the observed patterns and is
not intended to reflect the order of outages when the pattern is produced by the protection system. It aims to represent the statistics of the overall effect observed after all the protection mechanisms have acted.

The sampling procedure to match the key statistics of the patterns developed in section~\ref{statistics} 
is as follows:

(1) The assumed initial single line outage \poneone is chosen at random. The single-line outage can be caused by various factors such 
as bad weather or a fault. In system risk simulations, it is common to assume a uniform or length-dependent 
line outage probability that is applied to the lines in the network. 

(2) We determine how many lines are in the pattern according to the Zipf distribution fitting the data 
in Fig.~\ref{distnlinesinpatterns}. 
 A simple way to do this is to sample from the positive integers with weights given by the Zipf distribution. A more accurate and efficient method would be to use stratified sampling as in \cite{KellyPS24} with each positive integer up to a bound as a stratum. 
 The first seven numerical values of the Zipf distribution probabilities are shown in  Table~\ref{zipfvalues}.

\begin{table}[hbpt]
	\caption{Probabilities of number of lines in a pattern}
	\label{zipfvalues}
	\centering
	\setlength{\tabcolsep}{0.3em}
	\begin{tabular}{ cccccccccc }
	&1&2&3&4&5&6&7\\[0.2mm]
	\hline		\\[-2.5mm]
		\text{BPA} & 0.92911 & 0.05451 & 0.01038 & 0.00320 &
   0.00128 & 0.00061 & 0.00032  \\
		\text{NYISO} & 0.93336 & 0.05179 & 0.00954 & 0.00287 &
   0.00113 & 0.00053 & 0.00028 \\					        		\hline			
							\end{tabular}
\end{table}
(3) We describe {\sl where} an additional line is added
to an evolving pattern.

(3a) If another outaged line is added to the initial outaged line \poneone to form \ptwooneone, the new outaged line is attached at one of the two buses of \poneone. This is always possible since the network is connected. 
If several neighboring lines are available to be added, one of these neighboring lines is selected with equal probability.

(3b) 
Consider that the evolving pattern already has $\ge2$ lines, and another line in the network is to be added to the pattern. If there are lines in the network available to add at either a bus of degree 1 in the evolving pattern or at a bus of degree $\ge2$ in the evolving pattern,  then the line is added to a bus of degree 1 in the evolving pattern with probability $p_{1+}^{\rm }$, and added to a bus of degree $\ge2$  with probability $1-p_{1+}^{\rm }$.
However, for a modest fraction of evolving patterns, the network constrains which lines are available in the network to add to the pattern.
If there are only lines in the network available to add at a bus of degree 1 in the evolving pattern, then the line is added at a bus of degree 1 in the evolving pattern.
If there are only lines in the network available to add at a bus of degree $\ge2$ in the evolving pattern, then the line is added at a bus of degree $\ge2$ in the evolving pattern.
If the network allows several choices of lines to add to the evolving pattern in one of these ways, then select one of the choices with equal probability.
The calculation of $p_{1+}^{\rm }$ is explained  in section~\ref{fitinnetwork}. 

(4) For the BPA network, if any line in the pattern has multiple circuits in parallel, outage one of the other parallel circuits with probability $p_{\ge 2}^{\rm circuits}$.

\looseness=-1
Note that the objective is not to reproduce the observed patterns exactly but to be able to generate new patterns that are credible because they are statistically similar in key respects. 
If one wanted to use the exact patterns, then this could also be done directly from the specific historical patterns observed, but this lacks flexibility in the number and variety of samples, and since the larger patterns are rare,
the historical patterns are only a limited sample of the possibilities.
Our model is able to generate rare large patterns in new locations. There is a uniformity assumption that the same generative algorithm applies uniformly across the network, but this assumption is reasonable for this first generative model. Sometimes, it may be useful to consider simplified models that restrict the patterns generated. For example, the very simplest model adds a second adjacent line with probability 7\% to any given single line outage.

\begin{figure}[htb]
\centering
\includegraphics[width=0.86\linewidth]{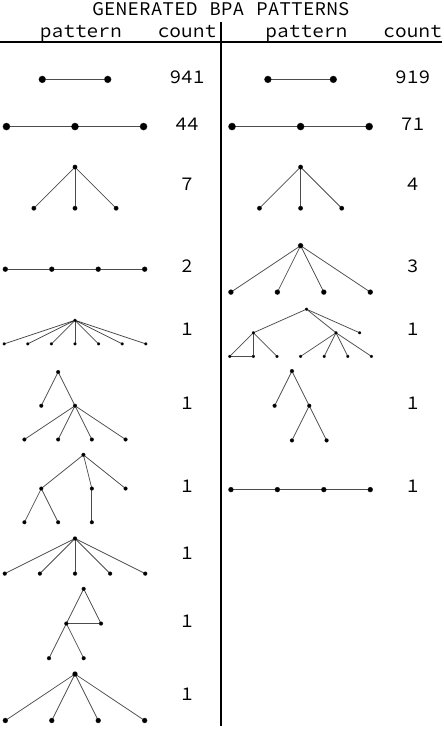}
\caption{\label{patternsgeneratedBPA}
Two samples of 1000 BPA generated patterns are indicated by the counts of their degree sequences and an example pattern for each degree sequence. In these two particular samples, because there is only one of the more complicated patterns generated, and the degree sequence determines the pattern for the simpler patterns, all the generated patterns are shown and the count of each degree sequence is the same as the count of each pattern.  Note that any multiple circuits outages in the patterns are not shown.}
\end{figure}

\begin{figure}[htb]
\centering
\includegraphics[width=0.86\linewidth]{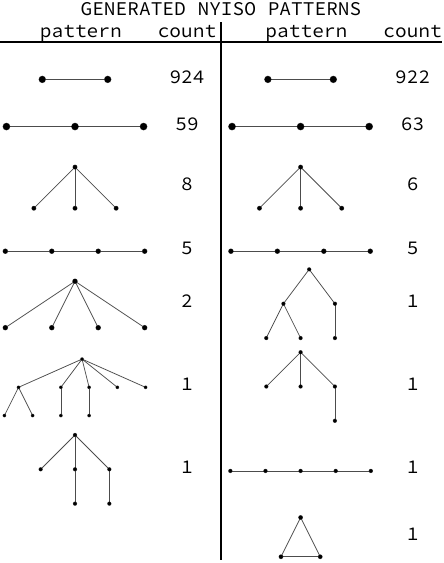}
\caption{\label{patternsgeneratedNYISO}
Two samples of 1000 NYISO generated patterns are indicated by the counts of their degree sequences and an example pattern for each degree sequence. In these two particular samples, because there is only one of the more complicated patterns generated, and the degree sequence determines the pattern for the simpler patterns, all the generated patterns are shown and the count of each degree sequence is the same as the count of each pattern. }
\end{figure}

Two samples of 1000 generated patterns for each transmission system are shown in Figs. \ref{patternsgeneratedBPA} and \ref{patternsgeneratedNYISO}.

Since the generative model randomly selects an  initial line and only adds some adjacent lines probabilistically, its execution is very fast and largely independent of network size. 

\section{Protection Event Propagation Slope Index}

The empirical distribution of the number of outaged lines in a pattern on a log-log plot is fit with a straight line in Fig.~\ref{distnlinesinpatterns} for both BPA and NYISO. 
The magnitude of the slope of the line is the parameter $s$ of the fitted Zipf distribution (\ref{zipf}). 
We suggest using $s$ as a Protection Event Propagation Slope Index (PEPSI). For example, the slope magnitude of the BPA fitted line in Fig.~\ref{distnlinesinpatterns} is given by $\mbox{PEPSI}=s=4.09$. 
For NYISO, $\mbox{PEPSI}=4.17$. Using the fitted line smooths the more erratic data points caused by the sparse data for the larger patterns.

A lower value of PEPSI indicates a shallower slope and an increased probability of outages propagating to form larger patterns on the fast time scale.
More quantitatively, we can use PEPSI to estimate the probabilities of large patterns. These probabilities are conditional on an initial automatic outage happening.
 Define a large pattern as having 4 or more outages
 (different cut-offs can be chosen).
 Then, substituting PEPSI for $s$ in (\ref{zipf}), we can compute the probability  
 \begin{align}
& p_{\rm large}=\notag\\&\sum_{k=4}^\infty {\rm P}[Z=k\,|\,s=\mbox{PEPSI}]
 =1-\sum_{k=1}^3 {\rm P}[Z=k\,|\,s=\mbox{PEPSI}]\notag
 \end{align} 
For example, for ${\rm PEPSI}=4.1$, $p_{\rm large}=0.0059$, 
and for ${\rm PEPSI}=3.8$, $p_{\rm large}=0.0094$.

PEPSI is analogous to the System Event Propagation Index (SEPSI): 
PEPSI measures how much outages propagate at the fast protection time scale, whereas SEPSI measures how much generations of outages propagate in the network at a slower time scale. SEPSI is described in 
\cite{DobsonArxiv18,EkishevaPESGM21}.

Conventional protection system reliability metrics generally describe the probabilities or annual rates of specific types of misoperation or operation for specific components or subsystems \cite{Anderson}.
For example, breaker failures are quantified in \cite{Bollen93,AndersPS06,GrantPhD25}.
The PEPSI metric is different in that it measures the overall spread of protection actions on an entire network for all causes of additional lines outaging. 

 \section{Accounting for how evolving patterns fit into the network}
 \label{fitinnetwork}

 Generating the patterns requires a value of $p_{1+}^{\rm }$, which is the probability of adding a line to a bus of degree 1 in the evolving pattern  when there is a choice available in the network between adding a line at a bus of degree 1 in the evolving pattern and adding a line at a bus of degree $\ge2$ in the evolving pattern.

 The patterns are generated by successively adding outaged lines to a randomly chosen initial outage. And whether a line will be added to buses of degree 1 in an evolving pattern with more than one line
should match the observed probability $p_{1+}^{\rm observed}$. 
However, for some of the evolving patterns that are unfavorably positioned in the network it can happen 
that
the network does not have any line available to attach to a bus of degree 1 in the pattern, or the network only has lines available to attach to a bus of degree 1 in the pattern. 
This section computes a value of $p_{1+}^{\rm }$ accounting for this effect.

We assume a known distribution of initial outages, which here is taken to be a uniform distribution.
We generate 1\,000\,000 patterns
assuming a value of $p_{1+}^{\rm }$ starting from the distribution of initial outages.
From these patterns, we empirically determine the probability $p_{1+}^{\rm generated}$, which is the probability of lines being attached at a bus of degree 1 in an evolving pattern with at least 2 lines. Then we repeat this calculation of $p_{1+}^{\rm generated}$, adjusting the value of $p_{1+}^{\rm }$ until $p_{1+}^{\rm generated}$ matches the corresponding quantity $p_{1+}^{\rm observed}$ observed in the utility data.
This calculation yields $p_{1+}^{\rm }=0.11$ for the BPA network and $p_{1+}^{\rm }=0.45$ for the NYISO network.

\section{Evaluating generative model results}

To evaluate the generative model, we introduce a distance metric that quantifies the difference between the observed data and the generative model results. Such distance metrics have been used to evaluate generative models in neuroscience, astrophysics, X-ray images, and other scientific settings \cite{Bischoff2024}. 
To construct this distance, we first define a distance between degree sequences of patterns based on the number of edges that need to be added or removed to convert one into another. Then we use the Wasserstein metric to define the distance between distributions of degree sequences of patterns. 
Given the Wasserstein distance, we can not only quantify how close the generated patterns are to the observed patterns but also statistically test whether the degree sequences of the generated and observed patterns can be considered to be samples from the same probability distribution. 

\subsection{Distance between degree sequences of patterns}

The distance $d(\delta_i, \delta_j)$ between the degree sequence $\delta_i$ of pattern $\pi_i$ and the degree sequence $\delta_j$ of pattern $\pi_j$ is defined as 
 the minimum number of line additions and subtractions to transform the degree sequence  $\delta_i$ into the degree sequence   $\delta_j$.
 This can be considered a restricted case of the graph edit distance or Hamming distance \cite{Donnat2018}. 
For instance, if the degree sequence of a pattern $\delta_i$ can be transformed to the degree sequence of $\delta_j$ by adding or removing one line, then $d(\delta_i, \delta_j)=1$.

\begin{figure}[htb]
\centering
\includegraphics[width=\columnwidth]{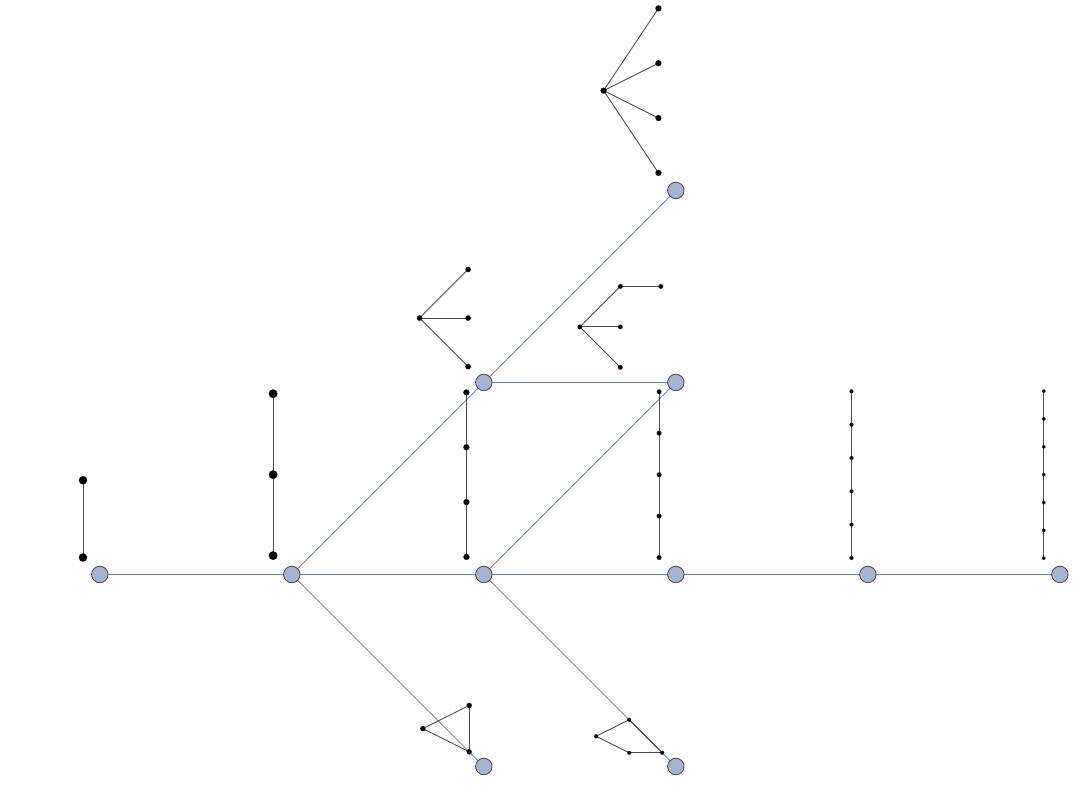}
\caption{Relation of the observed degree sequences in NYISO. Each degree sequence is labeled by an example pattern with that degree sequence (the degree sequence 5,2,1,1,1,1,1  requires two lines to be added to an observed degree sequence and is not shown). Degree sequences differing by the addition or subtraction of one line are joined by an edge. }
\label{relationgraphNYISO}
\end{figure}

To compute the pattern distance, it is convenient to form a graph with all the degree sequences of interest as nodes of the graph. The degree sequences nodes are connected by an edge of the graph if they are distance one apart.  Examples of such a graph, but only showing the observed degree sequences, are shown in Figs.~\ref{relationgraphBPA} and \ref{relationgraphNYISO}.

To form the graph, we start with the degree sequence 1,1 of $\poneone$ and successively add lines one at a time to obtain new degree sequences with each new degree sequence joined by an edge of the graph to its preceding degree sequence. 
All the possible line additions that maintain a connected graph with no multiple lines are considered. Since the edges of the graph join degree sequences that differ by one line, the minimum graph distance between two degree sequences is the distance between the degree sequences, and can be computed with Dijkstra's algorithm assuming the weight of every edge to be 1.

The detail of forming the graph is as follows: We 
regard the degree sequences as padded on the right with zeros 
to allow new buses to be added. Then adding a line changes the degree sequence by adding 1 to two of the bus degrees. 
The new line either connects two existing buses or joins a new bus to an existing bus.
All possible degree sequences produced by such line additions are considered, except that degree sequences corresponding to multiple lines joining the same two buses are excluded using the test for a simple graph from the Havel and Hakimi theorem \cite{hartsfield2013pearls}.

\subsection{Distance between distributions of degree sequences}

Next, to establish a distance between distributions of degree sequences, we use the Wasserstein metric or Earth Mover's Distance. By taking the union of all the degree sequences, and dividing the number of occurrences of each degree sequence by the total number of degree sequences, we can construct the empirical probability mass distribution for the degree sequences of both the observed and generated patterns 
\begin{align*}
P &= {(\delta_1, p_1), (\delta_2, p_2), \ldots, (\delta_n, p_n)} \\ \,
Q &= {(\delta_1, q_1), (\delta_2, q_2), \ldots, (\delta_n, q_n)}
\end{align*}
Here, $\delta_i$ is the degree sequences of pattern $i$, and $p_i$ or $q_i$ are the corresponding probabilities. The Wasserstein metric between these two discrete distributions computes the minimal transport plan that is required to convert one distribution into another based on the distance between degree sequences. For this particular case, the Wasserstein metric is the solution of a linear program 
\begin{align}
W(P, Q) = \min_{X} \sum_{i,j} x_{ij} \cdot d(\delta_i, \delta_j) 
\end{align}
subject to 
$\sum_j x_{ij} = p_i$ for all $i$,
$\sum_i x_{ij} = q_j$ for all $j$, and
$x_{ij} \geq 0$ for all $i, j$. Here $X=(x_{ij})$ is the transportation matrix or transportation plan.
There are several open source libraries that implement this calculation, and we use the Python Optimal Transport library (POT)  \cite{flamary2021pot}.

The Wasserstein distance between two distributions of degree sequences can be interpreted as the fraction of lines that have to be changed to convert one distribution of degree sequences into the other \cite{Bolt2022}. If the two distributions were obtained from the same number of degree sequences, then the  Wasserstein distance times the number of degree sequences can be   interpreted as 
number of lines that have to be changed to convert one set of degree sequences into the other. 

\subsection{Distance between the degree sequences of observed and generated patterns}

We evaluate the distance between the observed patterns and  the same number of samples of the generated patterns by computing the Wasserstein distance between the distributions of their degree sequences. Because  the generative model produces different sets of patterns every time it runs, there is variance in the evaluation of this distance. 
Thus, we perform this evaluation 1000 times and compute statistics of the result. 

For the BPA system, the mean distance between the generated and observed patterns is 0.009 while its standard deviation is 0.002. 
The Wasserstein distance of 0.009 corresponds to changing $0.009\times 11836=107$ lines to change the 11836 generated pattern degree sequences  into those observed. For the NYISO system,  the mean distance of the generated and observed data is 0.013 while its standard deviation is 0.003.
The Wasserstein distance of 0.013 corresponds to changing $0.013\times 7362=99$ lines to change the 7362 generated pattern degree sequences into those observed. We regard these distances for BPA and NYISO as satisfactorily small.

\subsection{Statistical testing of the observed and generated patterns}

We statistically test whether the generated patterns and the observed patterns are from the same underlying probability distribution using a permutation test \cite{good2013permutation} with the test statistic of distance between the distributions of the degree sequences of the observed and generated patterns.
Permutation tests are applied in \cite{sutherland2017generative} to test generative models, and 
 have been proposed to validate synthetic datasets in finance, healthcare, and other fields \cite{visani2022enabling}.

The generative model is used to produce 1000 sets of patterns, with each set containing the same number of patterns as those observed (11836 patterns for BPA and 7362 patterns for NYISO). A permutation test is run  on each of the 100 sets of patterns to test the null hypothesis that each set of patterns is from the same underlying distribution as the observed patterns. 
Each permutation test samples 10\,000 permutations.
The p-value for each test is the estimated probability that the distance of the generated set of patterns from the observed set of patterns exceeds the original unpermuted distance. We do not reject the hypothesis that the set of patterns is from the same underlying distribution if the p-value$\,\ge0.05$.

This testing procedure gives 1000 p-values for each transmission system. For BPA, the median p-value is 0.31, 97\% of the p-values exceed 0.05, and 99.9\% of  the p-values exceed 0.01, and for NYISO, the median p-value is 0.08, 69\% of the p-values exceed 0.05, and 93\% of the p-values exceed 0.01.  
These results indicate that in almost all cases for BPA and in most cases for NYISO, the samples from our generative model are indistinguishable from the observed data, supporting the case  that our model can often capture the underlying statistical structure of outage propagation.

The comparison between the generated and observed patterns is necessarily statistical  in nature. While the permutation test inherently can only fail to disprove (rather than prove) the hypothesis that the generated and observed patterns are drawn from the same statistical distribution, the permutation test results comparing the closeness of the degree sequences of the generated and observed patterns are encouraging. 
The degree sequence is a significant characteristic of a pattern and the distance between degree sequences incorporates the intuition that close patterns differ by changing only a few lines.
However, there could potentially be other degrees of freedom not constrained by the distances between degree sequences that could show some significant differences between the generated and observed patterns.

The deviations from the observed patterns in the generative model include the random aleatory variations when sampling from the generative model as well as
epistemic imperfections in the form of the generative model, 
which may include the limited number of statistical characteristics reproduced by the generative model and a uniformity assumption.
By using the data for the limited number of larger historical patterns to generate statistically similar patterns across the network, the generative model relies on an assumption that there is sufficient uniformity across the network in the statistics of fast failure propagation. 
This is a reasonable assumption for a transmission system in one specific geographic region that is designed and maintained by a single regional entity.

\section{Discussion and Conclusions}

We show how to extract network patterns in transmission line outages from utility outage data at the fast protection time scale and the large transmission system size scale.
The utility data include the bus names at each end of the line and the outage start times to the nearest minute. 
The outage patterns are the connected groups of lines that outage starting in the same minute. It is easy to note which outages start in the same minute to form the patterns.  Then the patterns can be located on the network.  The required data is systematically collected by all transmission utilities across North America and reported to NERC in the TADS Transmission Availability Data System. Similar data is collected world wide.
Utilities know their own network topology so that the outage patterns can be located on their network. Alternatively, as done in this paper, given unique bus names, it is straightforward to form the network topology directly from the utility outage data itself so that the patterns of outages can be located on the network \cite{DobsonPS16}.
We conclude that the network patterns can easily be found from data that is generally collected by transmission utilities worldwide. 

We extract the fast time scale outage patterns for two transmission systems. 
This appears to be the first publication of such data, and it is informative to see the multiple outage patterns that actually occurred together with their frequencies.
The patterns at this fast time scale are quite different than the patterns formed by  outages that occur at a slower time scale due to extreme weather and cascading \cite{DobsonPS16}. 
The outage patterns at the fast time scale are almost all connected, whereas the outage patterns at the slower time scale tend to be disconnected.
We do suggest that separately analyzing the fast protection time scale and the slower time scales in analyzing extreme weather and cascading is valuable. 
Examining real data is a sound basis for these analyses, and this paper contributes to this at the fast protection time scale.

We approximate the observed distribution of the number of outages in each pattern with a Zipf distribution. 
The heavy-tailed nature of the Zipf distribution has the practical implication that the larger patterns,  although rare, can be expected to occur. 
The Zipf distribution is characterized by its slope on a log-log plot, and the magnitude of this 
slope provides a Protection Event Propagation Slope Index that describes 
how much outages spread in the network at the fast timescale.

The patterns with $\ge$3 lines can be evolved by starting with two adjacent line outages and adding lines. 
We show how to estimate from the observed final patterns and the network structure the probability that a line was added to buses already attached to only one line. 
This probability controls whether the pattern evolves tending towards linear strings of lines or tending towards stars in which multiple lines are attached to one bus. This probability, together with the statistics of the number of lines in a pattern, are key statistics describing the observed patterns.

These key statistics can be sampled to generate representative sets of outage patterns consistent with the observed statistics.
Indeed, this amounts to a novel generative model of the protection system's overall effects at the transmission system scale. 
The generative model is inherently computationally fast, producing network patterns very quickly even in large networks.

\looseness=-1
The ability to generate representative outage patterns from any given starting line outage, caused, for instance, by weather or by equipment failure, is important because the historical outages alone will not cover the full range of credible possibilities when assessing the risk of future outages.
Since they are based on utility data, 
 the generated patterns incorporate the common cases of routine operation as well as the rare,
but more impactful complicated cases  (protection backup, substation design, stuck breaker, hidden failure, common mode etc.) in which outages quickly spread further in the network.
The patterns of outages on the network can be generated  according to their statistics without getting involved in the various mechanisms
for their cause and the 
formidable difficulties in practice of obtaining and representing
the details of the protection system and substations across an entire transmission system.
 It is best to apply the generative model to a transmission system with the parameters obtained from the detailed outage data for that transmission system. However, if the detailed outage data is not available, as for example in IEEE test systems or synthetic grids, the generative model could be applied with parameters typical of those observed in other systems. 
 The generative model is inherently computationally fast, producing network patterns very quickly even in large networks.

\looseness=-1
We show sample results of the generated patterns. 
To test these generated patterns quantitatively, 
we define distances between sets of patterns and show that the generated patterns are close to the observed patterns. Moreover, we consider whether the generated patterns are from a different probability distribution than the observed patterns when differences in their distance are evaluated.
A statistical permutation test shows that it is unlikely in most cases that the generated patterns are from a different underlying probability distribution than the observed patterns.

This paper is devoted to analyzing observed utility data and developing a new generative model of patterns of outages at the large transmission network scale and the fast protection system time scale. The required data is generally available to transmission utilities and the required computations are straightforward, fast, and scalable, so that finding the observed patterns and generating statistically similar patterns is widely applicable.  Promising future applications for the new generative model 
include contributing a practical statistical model of the effects of protection that can be included in Monte Carlo simulations assessing blackout risk at the transmission system scale, 
providing observed data to calibrate other  protection models, and 
generating enough synthetic data for rare protection events to train AI models.

\section*{ACKNOWLEDGEMENT}
\noindent
We gratefully thank Bonneville Power Administration and New York Independent System Operator for making publicly available the outage data that made this paper possible. The analysis and any conclusions are strictly those of the authors and not of BPA or NYISO.

 \vspace{0.3cm}
    \begin{center}
        \scriptsize \framebox{
            \parbox{2.5in}{
                Government License (will be removed at publication):
                The submitted manuscript has been created by UChicago Argonne, LLC,
                Operator of Argonne National Laboratory (``Argonne").  Argonne, a
                U.S. Department of Energy Office of Science laboratory, is operated
                under Contract No. DE-AC02-06CH11357.  The U.S. Government retains for
                itself, and others acting on its behalf, a paid-up nonexclusive,
                irrevocable worldwide license in said article to reproduce, prepare
                derivative works, distribute copies to the public, and perform
                publicly and display publicly, by or on behalf of the Government. The Department of Energy will provide public access to these results of federally sponsored research in accordance with the DOE Public Access Plan. http://energy.gov/downloads/doe-public-access-plan.
            }
        }
        \normalsize
    \end{center}


\begin{thebibliography}{99}

\bibitem{DobsonPMAPS18} I. Dobson, A. Flueck, S. Aquiles-Perez, S. Abhyankar, J. Qi, Towards incorporating protection and uncertainty into cascading failure simulation and analysis, Probability Methods Applied to Power Systems Conf., Boise, Idaho USA, June 2018.

\bibitem{FlueckPESGM20}  A.J. Flueck et al., 
Dynamics and protection in cascading outages, IEEE Power and Energy Society General Meeting, Montreal CA, August 2020.

\bibitem{Aragon2017} P. Aragón, V. Gómez, D. García, A. Kaltenbrunner,  Generative models of online discussion threads: state of the art and research challenges, Journal of Internet Services and Applications, vol. 8, Oct. 2017.  

\bibitem{ZhouPS24} K. Zhou, I. Dobson, Z. Wang, The most frequent N-k line outages occur in motifs that can improve contingency selection, IEEE Trans. Power Systems, vol. 39, no. 1, pp. 1785-1796, Jan. 2024.

\bibitem{ChenPS05} Q. Chen, J.D. McCalley, Identifying high risk N-k contingencies for online security assessment, IEEE Trans. Power Systems, vol. 20, no. 2, pp. 823-834, May 2005.

\bibitem{YangIREP07} F. Yang, S. Meliopoulos, J. Cokkinides,  Q.B. Dam, Bulk power system reliability assessment considering protection system hidden failures, Proc. IREP Symp., Aug. 2007, pp. 3408–3421.

\bibitem{JiangNAPS21}
Y. Jiang et al., Contingency probability estimation for risk-based planning studies using NERC’s outage data and standard TPL-001-4, in Proc. IEEE North American Power Symp., 2021, pp. 1–6.

\bibitem{CFTFPESGM08} IEEE PES CAMS Task Force on Understanding, Prediction, Mitigation and Restoration of Cascading Failures, Initial review of methods for cascading failure
analysis in electric power transmission systems, IEEE PES General Meeting, Pittsburgh, PA USA July 2008.

\bibitem{WGPS16} J. Bialek et al.,
Benchmarking and validation of cascading failure analysis tools, IEEE Trans. Power Systems, vol. 31, no. 6, Nov. 2016, pp. 4887-4900.

\bibitem{RiosPS02} M.A. Rios, D. S. Kirschen, D. Jayaweera, D.P. Nedic, R.N. Allan, Value of security: modeling time-dependent phenomena and weather conditions, IEEE Trans. Power Systems, vol. 17, no. 3, pp. 543-548, Aug. 2002.

\bibitem{ChenEPES05} J. Chen, J.S. Thorp, I. Dobson, Cascading dynamics and mitigation assessment in power system disturbances via a hidden failure model, Intl. J. Elect. Power \& Energy Sys., 2005, vol. 27, no. 4, pp. 318-326.

\bibitem{YuPS04} X. Yu, C. Singh, A practical approach for integrated power system vulnerability analysis with protection failures, IEEE Trans. Power Systems, vol. 19, no. 4, pp. 1811–1820, Nov. 2004.

\bibitem{CiapessoniSG16} E. Ciapessoni et al., 
Probabilistic risk-based security assessment of power systems considering incumbent threats and uncertainties, IEEE Trans. Smart Grid, vol. 7, no. 6, pp. 2890-2903, Nov. 2016.


\bibitem{AndersPS06} G.J. Anders, H. Maciejewski, B. Jesus, F. Remtulla, A comprehensive study of outage rates of air blast breakers, IEEE Trans. Power Systems, vol. 21, no. 1, pp. 202-210, Feb. 2006.

\bibitem{Bollen93} M.H.J. Bollen, Literature search for reliability data of components
in electric distribution networks, EUT Report 93-E-276, Eindhoven
University of Technology, Netherlands, Aug. 1993.

\bibitem{GrantPhD25} J. Grant, The impact of high-voltage circuit breaker condition on power system reliability, PhD thesis, Norwegian University of Science and Technology, 2025.

\bibitem{MittelstadtPMAPS04} W.A. Mittelstadt, D.P. Ferron, S.K. Agarwal, R.E. Baugh, Outage probability evaluation of lines sharing a common corridor,  Probabilistic Methods Applied to Power Systems, Ames, IA, USA, 2004, pp. 274-279.

\bibitem{KeelPESGM12} B. Keel, M. Papic, D. Tucker, Western Electricity Coordinating Council experience in the collection of transmission common-mode and dependent outages, 2012 IEEE PES General Meeting, San Diego, CA, USA, 2012.



\bibitem{PapicPESGM14} M. Papic et al., Effects of dependent and common mode outages on the reliability of bulk electric system - Part II: Outage data analysis, 2014 IEEE PES General Meeting, National Harbor, MD, 2014.

\bibitem{BillintonPESGM12} R. Billinton, Basic models and methodologies for common mode and dependent transmission outage events, 2012 IEEE PES General Meeting, San Diego, CA, USA, 2012.

\bibitem{PapicPS17} M. Papic et al., Research on common-mode and dependent (CMD) outage events in power systems: A review, IEEE Trans. Power Systems, vol. 32, no. 2, pp. 1528-1536, March 2017.



\bibitem{KellyEPSR20}  M.R. Kelly-Gorham, P.D.H. Hines, K. Zhou, I. Dobson, Using utility outage statistics to quantify improvements in bulk power system resilience,
Power Systems Computation Conf., Porto, Portugal, June 2020
and Electric Power Systems Research, vol 189, 106676, Dec. 2020.

\bibitem{KellyPS24}  M.R. Kelly-Gorham, P.D.H. Hines,  I. Dobson, Ranking the impact of interdependencies on power system resilience using stratified sampling of utility data, 
IEEE Trans. Power Systems, vol. 39, no. 1, Jan. 2024,
pp. 1251-1262.

\bibitem{ChengES22} B. Cheng, L. Nozick, I. Dobson, Investment planning for earthquake-resilient electric power systems considering cascading outages, Earthquake Spectra, vol. 38, no. 3, pp.1734-1760, 2022.

\bibitem{Leskovec2007} J. Leskovec et al., 
Cascading behavior in large blog graphs, April 2007. http://arxiv.org/abs/0704.2803

\bibitem{Gomez2011} V. Gómez,  H. Kappen, A. Kaltenbrunner, Modeling the structure and evolution of discussion cascades, Proc. 22nd ACM Conference Hypertext and Hypermedia, pp. 181-190, 2011. 

\bibitem{BPAwebsite} ``Bonneville power administration Transmission services Operations \& reliability Operations information” 2024. [Online]. Available: 
www.bpa.gov/energy-and-services/transmission/operations-information

\bibitem{NYISOwebsite} “New York independent system operator website,”  2024
[Online]. Available: http://mis.nyiso.com/public/P-54Blist.htm

\bibitem{CarringtonNAPS21}  N.K. Carrington, I. Dobson,  Z. Wang, Transmission grid outage statistics extracted from a web page logging outages in Northeast America, in Proc. IEEE North Amer. Power Symp., 2021, pp. 1–6.

\bibitem{DobsonPS16} I. Dobson, B.A. Carreras, D.E. Newman,  J.M. Reynolds-Barredo,
Obtaining statistics of cascading line outages spreading in an electric transmission network from standard utility data, IEEE Trans. Power Systems, vol. 31, no. 6, November 2016, pp. 4831-4841.

  \bibitem{DobsonPS12}
I.~Dobson, Estimating the propagation and extent of cascading line outages
  from utility data with a branching process, IEEE Trans. Power
  Systems, vol.~27, no.~4, pp.~2146-2155, 2012.

  \bibitem{StallAOT09} C.A. Stall, K.L. Cummins, E.P. Krider, J.A. Cramer, Detecting multiple ground contacts in cloud-to-ground lightning flashes, J. Atmospheric Oceanic Technology, vol. 26, 2009, pp. 2392–2402.

\bibitem{PetersonESS24} M. Peterson, The thunderstorms
with the greatest lightning densities on
Earth, Earth and Space Science, vol. 11,
e2023EA003304, 2024.
  
\bibitem{ClausetSIAMREVIEW09} A. Clauset, C.R. Shalizi, M.E.J. Newman,
Power-law distributions in empirical data,
SIAM Review, vol. 51, no. 4,
Nov. 2009,
pp.661-703.

\bibitem{DobsonArxiv18} I. Dobson, Finding a Zipf distribution and cascading propagation metric in utility line outage data, arXiv:1808.08434 [physics.soc-ph], 2018.

\bibitem{EkishevaPESGM21} S. Ekisheva, R. Rieder, J. Norris, M. Lauby, I. Dobson, Impact of extreme weather on North American transmission system outages, IEEE PES General Meeting, Washington DC USA, July 2021.


\bibitem{Anderson} P.M. Anderson et al., Power System Protection, 2nd edition, Wiley, Hoboken, 2022.

\bibitem{Bischoff2024} S. Bischoff et al.,  A practical guide to statistical distances for evaluating generative models in science, (arXiv,2024), https://arxiv.org/abs/2403.12636
 
\bibitem{Donnat2018} C. Donnat, S.  Holmes,  Tracking network dynamics: A survey using graph distances, Annals  Applied Statistics, vol. 12, June 2018. 

\bibitem{hartsfield2013pearls}
N. Hartsfield, G. Ringel, \textit{Pearls in Graph Theory: A Comprehensive Introduction},
Dover, NY 2013.

\bibitem{flamary2021pot}
R. Flamary et al., POT: Python Optimal Transport, Journal  Machine Learning Research, vol. 22, no. 78, pp. 1--8, 2021. [Online]. Available: \url{http://jmlr.org/papers/v22/20-451.html}

\bibitem{Bolt2022} G. Bolt, S. Lunagómez, C. Nemeth, Distances for comparing multisets and sequences, arXiv:2206.08858v1 [stat.ME] 17 Jun 2022.



\bibitem{good2013permutation} P. Good, {\sl Permutation tests: a practical guide to resampling methods for testing hypotheses}, Springer Science \& Business Media,
2013

\bibitem{sutherland2017generative}
D.J. Sutherland et al., 
Generative models and model criticism via optimized maximum mean discrepancy,
Intl. Conf.  Learning Representations, 2017.
\url{https://openreview.net/forum?id=HJWHIKqgl}

\bibitem{visani2022enabling} 
G.~Visani et al.,
Enabling synthetic data adoption in regulated domains, IEEE 9th Intl. Conf. Data Science \& Advanced Analytics, 2022.

\end{thebibliography}
\end{document}